\documentclass[11pt]{article} 
\usepackage{appb,epsf}

\newcommand{\beq}{\begin{eqnarray}}
\newcommand{\eeq}{\end{eqnarray}}
\newcommand{\ba}{\begin{array}}
\newcommand{\ea}{\end{array}}

\newcommand{\lsim}{\raisebox{-0.13cm}{~\shortstack{$<$ \\[-0.07cm] $\sim$}}~}

\def\plb#1#2#3{{\it Phys. Lett. }{\bf B#1~}(19#2)~#3}

\def\npb#1#2#3{{\it Nucl. Phys. }{\bf B#1~}(19#2)~#3}

\def\hepph#1{{\bf hep-ph}/#1}
\begin{document}
\setlength{\unitlength}{1cm}
\sloppy

\begin{flushright}
KA--TP--17--1997\\
{\tt hep-ph/9710497}\\
October 1997 \\
\end{flushright}

\vspace{1cm}

\centerline{\large \bf One--loop weak dipole moments of heavy fermions}

\def\thefootnote{\fnsymbol{footnote}}

\vspace{0.15cm}
\centerline{\large \bf in the MSSM\footnote{
       Presented at the XXI School of Theoretical Physics 
       ``Recent Progress in theory and phenomenology of 
       fundamental interactions", 
       Ustro\'n, Poland, September 19-24, 1997.}} 

\vspace{0.3cm}
\centerline{\sc Jos\'e I. Illana} 

\vspace{0.3cm}
\centerline{Institut f\"ur Theoretische Physik, Universit\"at Karlsruhe,}

\centerline{D--76128 Karlsruhe, FR Germany}

\begin{abstract}

The MSSM predictions at the one-loop level for the weak dipole moments of
the $\tau$ lepton and the $b$ quark are presented. The imaginary part of the
AWMDM is of the order of the SM contribution whereas the real part may 
be a factor 5 (20) larger for the $\tau$ ($b$) in the high $\tan\beta$ scenario,
still a factor five below the QCD contribution in the $b$ case. More 
interestingly, a contribution up to twelve orders of magnitude larger than
in the SM may be obtained, already at the one-loop level, for the WEDM in a 
MSSM with complex parameters.

\end{abstract}
\PACS{12.60.Jv, 13.40.Em, 14.60.Fg, 14.65.Fy}

\def\thefootnote{\arabic{footnote}}
\setcounter{footnote}{0}

\section{Introduction}

The investigation of the electric and magnetic dipole moments 
provides very accurate tests of the quantum structure of the Standard
Model (SM) and its possible extensions.
As a generalization of the electromagnetic dipole moments of fermions
(AMDM and EDM), one can define weak dipole moments (WDMs), corresponding to 
couplings with a $Z$ boson instead of a photon.
The most general Lorentz structure of the vertex function that 
couples a $Z$ boson and two on-shell fermions (with outgoing momenta
$q$ and $\bar{q}$) can be written in term of form factors $F_i(s\equiv
(q+\bar{q})^2$) as 
\beq
\Gamma^{Zff}_\mu&=&ie\Bigg\{\gamma_\mu\left[\left(
F_{\rm V}-\frac{v_f}{2s_Wc_W}\right)-\left(F_{\rm A}-\frac{a_f}{2s_Wc_W}\right)
\gamma_5\right] \nonumber \\ & & +(q-\bar{q})_\mu[F_{\rm M}+F_{\rm E}\gamma_5] 
-(q+\bar{q})_\mu[F_{\rm S}+F_{\rm P}\gamma_5]\Bigg\}\ ,
\label{vertex}
\eeq
where $v_f\equiv(I^f_3-2s^2_WQ_f)$, $a_f\equiv I^f_3$.
The form factors $F_{\rm M}$ and $F_{\rm E}$ are related to the anomalous
weak magnetic and electric dipole moments of the fermion $f$ with mass $m_f$ 
as follows:
\beq
{\rm AWMDM}\equiv a^W_f&=&-2m_f\ F_{\rm M}(M^2_Z) \ , \nonumber\\
{\rm WEDM}\equiv d^W_f&=&ie\ F_{\rm E}(M^2_Z) \ ,\nonumber
\eeq
The $F_{\rm M}$ ($F_{\rm E}$) form factors are the coefficients 
of the {\em chirality-flipping} term of the CP-conserving (CP-violating) 
effective Lagrangian describing $Z$-fermion couplings. 

\begin{figure}[htb]
\begin{center}
\begin{picture}(10,4.7)
\epsfxsize=11cm
\put(-0.5,-0.3){\epsfbox{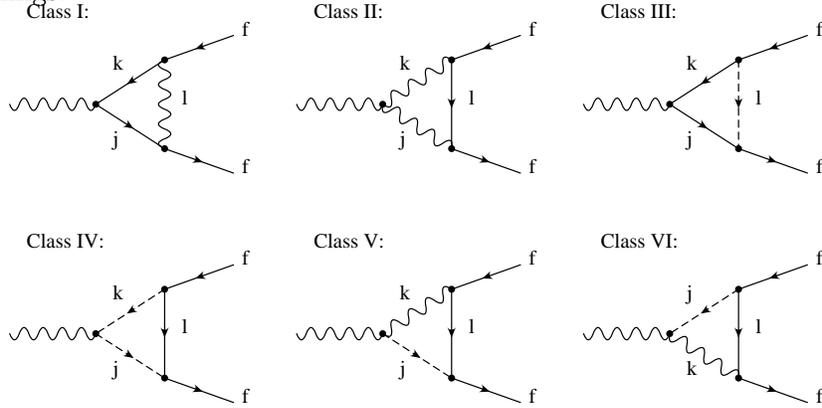}}
\end{picture}
\end{center}
\caption{The one-loop topologies for the $Zff$ vertex.
\label{fig}}
\end{figure}

In a renormalizable theory there is no contribution to the A(W)MDM or the
(W)EDM at tree level, as they correspond to operators of dimension 5. At
one loop, one can classify all the possible diagrams in six topologies or 
classes of triangle graphs \cite{topol} (Fig.~\ref{fig}). Analytical expressions
for these moments have been obtained \cite{WMDM,WEDM} in the 't Hooft-Feynman 
gauge to one loop. 
The global result, adding all the diagrams, is gauge independent, as it is
defined at $s=M^2_Z$.
The contribution of every class of diagrams is expressed in terms of 
standard three-point one-loop integrals and generic couplings, which allow for 
the implementation of different theories. Every term in the expressions
is proportional to some fermion mass, consistently with the chirality flipping
character of the dipoles. 
This  means that heavier fermions are expected to have larger WDMs. For an
on-shell $Z$ boson
the $b$ quark and $\tau$ lepton are the most promising candidates.
Unlike the electromagnetic dipole moments, defined
at $s=0$, the WDMs can be complex due to the possibility of pair-produce 
particles below the $M_Z$ threshold. 
Here the predictions for the MSSM are presented and compared to the SM ones.

\pagebreak
\section{Anomalous Weak Magnetic Dipole Moments}

The contribution to the AWMDM, for the $\tau$ lepton and for heavy quarks, 
at the one-loop level has been calculated by Bernab\'eu {\em et al.} in the
SM \cite{ber}. There the only free parameter is the SM Higgs
boson mass $M_{H^0}$ whose value does not significantly affect the
result\footnote{An opposite global sign is quoted here for comparison, due to 
our different conventions.} for 
$a^W_\tau=(2.10+0.61\ i)\times10^{-6}$ but it is more important for
the real part of $a^W_b=[(1.1;2.0;2.4)-0.2\ i]\times10^{-6}$, with
$M_{H^0}=M_Z,$ $2M_Z,$ $3M_Z$  respectively. Including the QCD contribution
(a gluon exchange in class I diagrams) in the case of the $b$ quark
dramatically enlarges the result to $a^W_b=(-2.96+1.56\ i)\times10^{-4}$.

In the MSSM the Higgs sector is different. It consists of a constrained 2HDM 
mainly controlled by the pseudoscalar Higgs boson mass $M_A$, the $\mu$ 
parameter and the ratio of VEVs $\tan\beta$. Besides, several  
soft-susy-breaking terms must be introduced. A simplified set is given by the 
assumption of universal scalar mass terms $m_{\tilde{q}}$ for squarks, and
$m_{\tilde{l}}$ for sleptons, trilinear terms 
$A_\tau$, $A_b$ and $A_t$ (for the third family) and gaugino mass terms related
by the GUT constraint: $\alpha M_3=\alpha_s s^2_W M_2=3/5\ \alpha_s c^2_W M_1$.
In Ref.~\cite{WMDM} a complete scan of the susy parameter space is performed.
The results for the AWMDM are summarized below.

The Higgs sector can provide the only contribution to the imaginary part,
of the order of the SM contribution, assuming the present experimental limits 
on the masses of the superpartners and MSSM Higgs bosons (except for still 
possible light neutralinos). The real part is typically
negative and not very large: $-$Re$\lsim0.3\times10^{-6}$ for the $\tau$ and
$-$Re$\lsim10^{-6}$ for the $b$.

The neutralino contribution to the real part is also small, and has opposite
sign than $\mu$ in most of the parameter space: $|$Re$(a^W_\tau)|\lsim0.02(0.4)
\times 10^{-6}$ and $|$Re$(a^W_b)|\lsim0.2(12)\times10^{-6}$ for $\tan\beta=1.6
(50)$ respectively.

The chargino contribution is the dominant one being real and with the
same sign as $\mu$: $|$Re$(a^W_\tau)|\lsim0.2(7)\times
10^{-6}$ and $|$Re$(a^W_b)|\lsim1(30)\times10^{-6}$.

The gluinos compete in importance with the charginos for the $b$ and their
contribution is always negative: $-$Re$\lsim2(40)\times10^{-6}$.

The sum of the MSSM contributions can amount to $|$Re$(a^W_\tau)|\lsim 0.5(7)
\times10^{-6}$  and $|$Re$(a^W_b)|\lsim2(50)\times 10^{-6}$
for not so extreme and not excluded regions of the susy parameter space. 
Decoupling is observed for large values of the parameters.

\newpage
\section{Weak Electric Dipole Moments}

In the SM there is only one source of CP violation, the $\delta_{\rm CKM}$ 
phase of the Cabibbo-Kobayashi-Maskawa (CKM) mixing matrix for quarks.\footnote{
The QCD lagrangian includes an additional source of CP violation, the
$\theta_{\rm QCD}$, but it will not be considered here.} 
The only place where CP violation has currently been measured, the neutral 
$K$ system, fixes the value of this phase but does not constitute itself a 
test for the origin of CP violation. Many extensions of the SM contain
new CP violating phases, in particular, the supersymmetric models.
The most significant effect of the CP violating phases in the phenomenology is 
their contribution to electric dipole moments. Their investigation may shed
some light on the intriguing problem of CP violation.

The SM contribution to the (W)EDM comes at three loops and hence it is as
small as $\sim e G_F m_f \alpha^2\alpha_s J/(4\pi)^5$, with $J\equiv
c_1c_2c_3s^2_1s_2s_3s_\delta$ (an invariant under reparametrizations of the
CKM matrix). That is the (W)EDM$\sim3\times10^{-34}\ (10^{-33})\ e$cm for the 
$\tau$ ($b$).

In the MSSM new physical phases, provided by the soft-breaking terms, are 
brought into the game and their effects manifest themselves
already at the one-loop level. For simplicity, we restrict to 
generation-diagonal trilinear soft-breaking terms to prevent from FCNC.
In our analysis we do not make any additional assumption, except
for the unification of the soft-breaking gaugino masses at the GUT scale. 
However we do not assume unification of the scalar mass parameters or trilinear 
mass parameters.
In such a constrained framework the following susy parameters can be complex:
the $\mu$ parameter, the gaugino masses, the bilinear mixing mass parameter 
$m^2_{12}$ and the trilinear soft-susy-breaking parameters.
Not all of these phases are physical. Namely, the MSSM has two additional U(1) 
symmetries for vanishing $\mu$ and soft-breaking terms: the Peccei-Quinn and 
the R-symmetry. For non-vanishing  $\mu$ and soft-breaking terms these
symmetries can be used to absorb two of the phases by redefinition 
of the fields \cite{relax}. In addition, the GUT constraint leads to only one 
common phase for the gaugino mass terms. After these considerations, our choice
of CP violating physical phases is:\footnote{
The (common) phase of the complex gaugino mass terms as well as the phase of
$m^2_{12}$ are absorbed. With this choice, the susy preserving $\mu$ parameter 
remains complex.} 
$\varphi_\mu\equiv{\rm arg}(\mu)$, 
$\varphi_{\tilde{f}}\equiv{\rm arg}(m^f_{\rm LR})$ ($f=\tau,\ t,\ b$)
with $m^t_{\rm LR}\equiv A_t-\mu^*\cot\beta$ and 
$m^{\tau,b}_{\rm LR}\equiv A_{\tau,b}-\mu^*\tan\beta$. The scan of the
the parameter space leads to numerical results of the same order as the
AWMDM expressed in {\em magnetons} $\mu_f$ for the values of the intervening CP
violating phases that maximize the effect (1 $\mu_f\equiv e/2m_f=
1.7(0.7)\times10^{-15}\ e$cm for $\tau$ ($b$), respectively). The expressions
contain only imaginary parts of combinations of couplings \cite{WEDM}.

The MSSM Higgs sector is CP conserving and hence it does not contribute to
the (W)EDM.

The diagrams with neutralinos involve $\varphi_\mu$ and $\varphi_{\tilde{\tau}}$
($\varphi_{\tilde{b}}$) for the $\tau$ ($b$) case and its contribution is
maximal for these phases being $\pi/2$.

The chargino diagrams only involve $\varphi_\mu$ in the $\tau$ case (there is
no scalar neutrino mixing) and also $\varphi_{\tilde{t}}$ for the $b$.
The former contribution is enhanced for $\varphi_\mu=\pi/2$ and the two
phases conspire in the latter to yield a maximum effect for $\varphi_\mu=\pi/2$
and $\varphi_{\tilde{t}}=\pi$.

The gluinos contribute maximally to the $b$ WEDM for
$\varphi_{\tilde{b}}=\pi/2$.

The maximal global pure supersymmetric contribution amounts to 
$|$Re$(d^W_\tau)|\lsim 0.3(12)\times10^{-21}\ e$cm, 
$|$Re$(d^W_b)|\lsim 1.4(35)\times10^{-21}\ e$cm.

\vspace{0cm minus 0.3cm}
\section{Conclusions}

The MSSM predictions for the WDMs of the $\tau$ lepton and the $b$ quark are 
the following: 

$\bullet$
The real part of the AWMDM can reach values 5 (20) times larger than the
electroweak SM predictions for the $\tau$ lepton ($b$ quark) in the high
$\tan\beta$ scenario, still a factor five below the QCD contribution in the 
$b$ case. 

$\bullet$
In a generalized MSSM with complex parameters and generation-diagonal trilinear
soft-susy-breaking terms a contribution to the WEDM is possible already at
the one-loop level. It is governed by two physical phases ($\varphi_\mu$, 
$\varphi_{\tilde{\tau}}$) in the $\tau$ case and three physical phases
($\varphi_\mu$, $\varphi_{\tilde{b}}$ and $\varphi_{\tilde{t}}$) in the $b$
case. The WEDM can be as much as twelve orders of magnitude larger than
the SM prediction.

\vspace{0.2cm}
\centerline{\em Acknowledgements}

It is a pleasure to acknowledge W. Hollik, S. Rigolin and D. St\"ockinger
for a very fruitful collaboration on the subject.
I am also grateful to the Organizing Committee for the excellent organization
and the nice atmosphere of the School. 
This work has been supported by the Fundaci\'on Ram\'on Areces and partially 
by the Spanish CICYT under contract AEN96-1672.


\vspace{0cm minus.3cm}


\end{document}